\documentstyle{CUPconf}

\def\etal{\nobreak\mbox{\it et~al.~}}

\begin{document}

\title[Chemical Evolution of the Galaxy]{Chemical Evolution of the Galaxy}
\author[Tosi ]{%
M\ls O\ls N\ls I\ls C\ls A\ns
T\ls O\ls S\ls I\ns }

\affiliation{Osservatorio Astronomico di Bologna, Via Zamboni 33, I-40126,
Bologna, Italy}
\maketitle

\begin{abstract}
{\it Standard} models for the chemical evolution of the Galaxy are reviewed
with
particular emphasis on the history of the abundance gradients in the disk. The
effects on the disk structure and metallicity of gas accretion
are discussed, showing that a significant fraction of
the current disk mass has been accreted in the last Gyrs and that the chemical
abundances of the infalling gas can be non primordial but should not exceed
$\sim$0.3 Z$_{\odot}$.
The distributions with time and with galactocentric distance of chemical
elements are discussed, comparing the observational data with the corresponding
theoretical predictions by {\it standard} models, which reproduce very well the
ISM abundances at various epochs, but not equally well all the features
derived from observations of old stellar objects.
\end{abstract}

\section[]{Introduction}
In the last few years a new generation of models for the chemical evolution
of the Galaxy has started to appear, in which also the dynamics of the system
is taken into account (e.g. Sommer-Larsen \& Yoshii 1990, Chamcham
\& Tayler 1994, Hensler, this volume, and references therein).
This new class of models can provide a complete
scenario for the evolution of the Milky Way but is still in a rather
preliminary phase. The aim of this
presentation is then to review the current state of {\it standard}
models for the chemical evolution of the galactic disk, with particular
emphasis on the effect of gas accretion on the element abundances and
gradients.

These models are quite successful in accounting for the large scale - long
term phenomena taking place in the Galaxy, and reproduce its major observed
features, such as the age-metallicity relation and the G-dwarf distribution
in the solar neighbourhood, the elemental and isotopic abundances and ratios,
the present star formation rate, gas and total mass densities. To avoid
misleading conclusions, it is however necessary to test the models
by comparing their predictions not only with the observational constraints
derived for the solar neighbourhood but also with the data relative to other
galactic regions; first of all because the solar neighbourhood is not
representative of the whole disk, and secondly because the distribution with
galactocentric distance of several quantities provides important information
on the history of the Milky Way.

The current rate of star formation (SFR) represents an excellent example of the
importance of modeling the whole disk. One of the most popular approximations
for the SFR is a linear proportionality with the gas density, and some authors
consider it not only a simple and intuitive law, but also a realistic one
because it can reproduce several properties observed in the solar neighbourhood
objects. Since the observed radial distribution of the gas in the disk is
rather flat (see the shaded area in the bottom panel of Fig.2),
this approximation inevitably implies a flat radial distribution of
the present SFR. However, Lacey \& Fall (1985) have shown that the current
SFR in the disk, derived from a large sample of young objects (pulsars,
O-stars,
etc.), is actually a steep decreasing function of the galactocentric distance.
Thus, the radial distribution predicted by models assuming a SFR linearly
proportional to the gas density is totally inconsistent with the observed one.
On the contrary, models assuming the SFR proportional to both the gas and
the total mass densities (e.g. Tosi 1982 and 1988, Matteucci \& Fran\c cois
1989,
hereinafter MF)
are in agreement with the observed trend, thanks to the steep decrease of
the total mass density with galactic radius.

\section[]{Gas infall and abundance gradients}

The idea of a long-lasting infall of metal poor gas on the galactic disk
was first suggested by Larson (1972) and Lynden-Bell (1975) to solve several
inconsistencies of the first simple models: a too rapid gas consumption that
prevented to reproduce the amount of gas currently observed in the disk,
the unlikelihood of a complete collapse of the whole protogalactic
halo in a few 10$^8$ yr, and the existence of very few metal poor, long
living stars in the solar neighbourhood, compared to the relatively large
predicted percentage (the so called G-dwarf problem). Since then,
gas accretion has turned out to be necessary to explain most of the
characteristics of our disk, and all the chemical evolution models (with or
without dynamics) in better agreement with the largest set of observational
constraints assume a significant amount of gas infall throughout the
disk lifetime.

\subsection[]{Interstellar medium abundances}

One of the most evident effects of gas infall on galactic evolution concerns
the absolute value and the radial distribution of the element abundances.
If no gas accretion is assumed after the disk formation, all chemical evolution
models with reasonable SFR and initial mass function (IMF) predict too large
present abundances and/or inconsistent radial distributions. This problem is
apparent in the top panel of Fig.1 where the distribution of the current
oxygen abundances predicted by models with exponentially decreasing SFR and
Tinsley's (1980) IMF is compared with that derived by Peimbert (1979) and
Shaver \etal (1983) from HII regions observations. The models divide the disk
in concentric rings, 1 kpc wide, and assume the sun at 8 kpc. Gas motion can
be allowed between consecutive rings, whereas stars are assumed to die in the
same region where they were born. If the galactocentric distances of the
HII regions are properly rescaled assuming the sun at R=8 kpc,
the observational oxygen gradient is
$\Delta$log(O/H)/$\Delta$R = -0.103 dex kpc$^{-1}$. The long-dashed line
corresponds to a model with no infall after the disk formation 13 Gyr ago: it
is too flat and lies above the observed range of oxygen abundances. The solid
line, instead, fits very well the data and corresponds to a model with SFR
e-folding time of 15 Gyr and constant infall of primordial gas with density
rate F = 4 10$^{-3}$ M$_{\odot}$kpc$^{-2}$yr$^{-1}$ all over the disk.
This uniform density rate implies a larger mass of metal free gas infalling
in the outer than in the inner rings and favours the development of negative
metallicity gradients as steep as observed.
This model reproduces the most important features of the Milky Way and from
now on it will be referred to as the {\it reference} model. It must be
emphasized, however, that other combinations of the SFR and infall
parameters may lead to a similarly good agreement with the data, as
shown in Fig.1 by the short-dashed line, corresponding to a model assuming a
shorter e-folding time for the SFR (5 Gyr) and a lower infall density rate
(F = 2 10$^{-3}$ M$_{\odot}$kpc$^{-2}$yr$^{-1}$).
What is generally found is that models in better agreement with the
observational constraints assume SFR e-folding times in the range 5-15 Gyr
and e-folding times for the gas accretion rate longer than for the SFR.
This requirement
is not unrealistic, if we consider that according to Sofue's (1994) models the
Magellanic Stream is regularly supplying the Milky Way with gas since 10 Gyr.

\begin{figure}
  \vspace{9truecm}
  \caption{Present distribution of the oxygen abundance with galactocentric
  distance, as derived from HII region observations (dots) and theoretical
  models. The average observational uncertainty is shown in the bottom left
  corner. Top panel: models with primordial infall (see text for details).
  Bottom panel: models with metal enriched infall.}
\end{figure}

If the mass of infalling gas is assumed to increase inwards rather than
outwards the model predictions are much less satisfactory. The dotted line
corresponds to a model with infall rate proportional to the total mass of
each ring, which is then increasing toward the center. As a consequence,
there is a larger dilution of the inner interstellar medium (ISM) resulting
in a flat abundance distribution inconsistent with that derived from
HII regions.

The above results refer to infalling gas with primordial chemical composition,
which would be available at best in the intergalactic medium or in the early
halo. If the gas originates from regions already polluted by stellar
nucleosynthesis, such as the current halo or the Magellanic Stream,
it has most probably a non negligible metal content. The intermediate solid
line in the bottom panel of Fig.1 shows that if the
metallicity of the infalling gas is 0.5 Z$_{\odot}$ the predictions
of the {\it reference} model are at the upper edge of the observed
distribution.
The top solid line shows that if the infall metallicity is solar the resulting
oxygen abundance is definitely outside the observational range. From a large
variety of models, Tosi (1988b) has found
that to allow for a good agreement with the data the infall metallicity
should not exceed 0.3 Z$_{\odot}$. The same result has been obtained by
MF with different models and different assumptions on the relative abundances
of the infalling elements. This limit is perfectly consistent with
the metal content attributable to both the galactic halo and the Magellanic
Clouds.

Depending on the model parameters, the present infall rate for the whole
disk ranges between 0.3 and 1.8 M$_{\odot}$yr$^{-1}$. The lower limit of this
range is in agreement with the amount inferred from Very High Velocity Clouds
which, with the Magellanic Stream, are the most reliable observational
evidence for this phenomenon. If a fraction of High Velocity Clouds
could be considered of non disk origin as well (see Danly, this volume),
the amount of infalling gas observationally detected would cover all the
theoretical range.

\begin{figure}
  \vspace{9truecm}
  \caption{Radial distribution at three different epochs of quantities
   predicted by the {\it reference} model. Top panel: SF and infall rates;
   bottom panel: gas and total mass. The shaded area corresponds to the
   observed gas mass range as published by Lacey \& Fall (1985)}
\end{figure}

The metallicity gradients predicted by chemical evolution models depend on
the ratio between the SFR and the interstellar and infall gas predicted at
each epoch and at each galactocentric distance. The top panel of Fig.2 shows
the radial distribution of the SFR and infall rate resulting from the
{\it reference} model at three different epochs: the dotted line corresponds
to the epoch of disk formation (assumed to be 13 Gyr ago), the dashed line
to the epoch of sun formation (8.5 Gyr later), and the solid line to the
present. Since the infall rate is assumed to be constant,
only one line appears in the
figure. The bottom panel of Fig.2 displays the radial distributions of the
gas and total mass at the same three epochs. The disk is supposed
to evolve from an initial configuration of pure gas with radially decreasing
mass (dotted line). The initial SFR is radially decreasing as well, so
that in inner regions there is more astration and therefore larger stellar
production of metals. However, the amount of ISM gas which must be polluted
by these metals is much larger in the inner regions and therefore
the efficiency of the ISM enrichment is quite modest. Thus, at early epochs
the predicted abundance gradients are either flat or even positive, depending
on the model parameters
(see also Moll\'a \etal 1990). After several Gyr, the situation changes
significantly, because the larger astration of the inner regions leads to
a higher gas consumption which is not totally compensated by infall
since the gas accretion is assumed to be increasing outwards. Thus, at a
certain time (see for instance the dashed line corresponding to the situation
4.5 Gyr ago) the larger SFR in the inner regions corresponds to a higher
efficiency in the metal enrichment of the medium and negative abundance
gradients start to develop. Since then, and as long as the star formation
activity remains a decreasing function of the galactocentric distance,
the slope of the gradients keeps steepening, because the gas radial
distribution
becomes increasingly flat, the infall dilution is more efficient outside, and
the metal enrichment inside. As shown in Fig.2, the disk of the Galaxy is
currently in this phase with a very flat radial distribution of the gas mass,
a radially decreasing SFR and a SFR/infall rate ratio progressively smaller
for increasing R and equal to 1 at 9-10 kpc (see also Wilson \& Matteucci
1992).

A steepening with time of the abundance gradients is predicted by most
of the models (with or without dynamics) which are able to reproduce the
observational features of the Galaxy (e.g. MF, Chamcham \& Tayler 1994,
Koppen 1994) despite the rather different model characteristics (see, however,
Ferrini \etal 1994 for different predictions). It is
then important to verify if this predicted trend is indeed
consistent with the available observational constraints. Since the gradient
derived from data on HII regions is representative of the situation in
the current ISM, to test the model predictions older objects must be
examined as well.

\begin{figure}
  \vspace{9truecm}
  \caption{Top panel: Radial distribution of the oxygen abundances 3 Gyr ago
   derived from the {\it reference} model (solid line) and from observations
   of PNeII (dots). The data are from PP and the dotted line represent their
   best fit. Bottom panel: same, but for the He abundance.}
\end{figure}

Planetary Nebulae, specially of Peimbert's (Peimbert \& Torres-Peimbert 1983)
type II (PNeII), are also
very good indicators of the ISM metallicity. PNeII have stellar progenitors
with lifetimes in the range 1-5 Gyr and therefore represent the ISM
conditions around 3 Gyr ago. Two recent and extensive studies (Pasquali
\& Perinotto 1993, hereinafter PP, and Maciel \& Koppen 1994) show that the
abundance gradients derived for several elements in PNeII are systematically
flatter than the corresponding gradients derived from HII regions. For
instance,
the oxygen gradient derived by PP is
$\Delta$log(O/H)/$\Delta$R = -0.03$\pm$0.01 dex kpc$^{-1}$ and that derived
by Maciel \& Koppen is -0.07$\pm$0.01. The latter authors have also found
hints of increasing slopes of the gradients with decreasing age of the PNe
(i.e. from type III to type I), in agreement with the model predictions.
Fig.3 shows the helium (bottom) and oxygen (top) abundances as derived by
PP from PNeII and the corresponding predictions of the {\it reference} model.
The agreement between the model solid line and the empirical best fit to the
data (dotted line) is excellent. This confirms that in the last few billion
years the slope of the abundance gradients in the ISM has actually steepened.

No other gaseous indicators are available to check whether the gradients were
increasingly flatter at earlier epochs. Stars and stellar clusters of
whatever age are instead visible in a fairly large range of distances
and can therefore indicate what was the earlier scenario.

\subsection[]{Stellar abundances}

As far as single stars are concerned, the situation is unfortunately rather
confuse. Lewis \& Freeman (1989) found no significant metallicity gradient
in a sample of 600 old K-giants, but more recently Edvardsson \etal (1993)
have argued that the radial metallicity distribution derived from
a sample of 189 F and G-dwarfs is similar to that derived from HII
regions. The major result of this accurate and extensive work is the
scatter on the derived abundances which turns out to be much larger than
the observational uncertainties and should then be considered an intrinsic
feature of the analysed stellar population. Edvardsson \etal therefore
avoided to formally derive the slopes of the abundance gradients, but one
can obtain them from their table 14 where the analysed stars are divided in
different groups according to their age and galactocentric distance and the
average [Fe/H] of each group is given. Despite the poorness of the sample
in the older and more distant bins, and the corresponding weakness of the
statistics, it is interesting to point out that the resulting formal slopes
get flatter for increasing age (i.e. toward earlier epochs) and that the
oldest bin even shows a positive gradient (derived from two single points,
however !), thus giving some further support to the predictions of the
chemical evolution models.

\begin{figure}
  \vspace{6truecm}
  \caption{Age-metallicity distribution for $\alpha$ elements as derived from
  the {\it reference}   model and from Edvardsson \etal (1993) data. Filled
  circles and dashed curve refer to stars in rings with 4$\leq$R$\leq$7 kpc,
  open circles and solid curve to stars with 7$\leq$R$\leq$9 kpc, crosses and
  dotted line to stars with 9$\leq$R$\leq$11 kpc. Age is in Gyr.}
\end{figure}

Another interesting feature of the Edvardsson \etal data is the different
distribution of metallicity with age for stars at different galactic locations.
As pointed out by Pagel (1994) and shown in Fig.4, if one divides their stars
into three groups according to their mean galactocentric distances (inner
objects, solar ring objects, and outer ones), one finds
that the outer stars show a much flatter age-metallicity distribution,
with average abundances in the last ten billion years (i.e. over most of the
disk lifetime) systematically lower that those of the other objects. As already
mentioned above, the data show a large intrinsic scatter in the derived
metallicities of stars of any age and this scatter cannot be directly
reproduced by standard chemical evolution models like the {\it reference}
one which assumes both the SFR and the gas accretion in a sort of steady-state.
However, the age-metallicity relations predicted for the three ranges of
galactocentric distances
are consistent with each of the corresponding average empirical
distributions. Notice that the relations predicted for the solar (solid
line) and the outer (dotted line) rings flatten off at recent epochs, whereas
the relation predicted for the inner ring (dashed line) keeps increasing
up to the present time, as already shown by MF. Besides, Fran\c cois \&
Matteucci (1993) have argued that even the spread of the observed
age-metallicity distribution can be accounted for by {\it standard} models,
once the different birthplaces of the sample stars are considered.

The major problem of single stars analyses is the uncertainty in the derived
R, age and metallicity of objects beyond a limited distance from the sun
as confirmed by the Edvardsson's \etal survey. From this point of view,
open clusters are in principle safer indicators. There are, however, several
problems affecting also the determination of the cluster parameters, such as
the non homogeneity of most age estimates and the uncertainty on the cluster
original birthplace, the cluster disruption due to disk friction which can
alter the original distributions, etc. (see, however, Carraro \& Chiosi 1994).

Several years ago, young open clusters have been suggested to indicate steeper
abundance gradients than old open clusters (Mayor 1976, Panagia \& Tosi 1981).
However, more recent and extensive studies (Friel \& Janes 1993, Thogersen
\etal 1993) do not seem to support this hypothesis. The metallicity gradient
derived by Janes and collaborators is
$\Delta$[Fe/H]/$\Delta$R = -0.09$\pm$0.02 dex kpc$^{-1}$
for the whole sample of clusters of any age, and does not seem to depend on
the cluster age. Bearing in mind that for field stars in the disk oxygen has
been empirically found to follow the relation [O/Fe]$\simeq-$0.3[Fe/H] (e.g.
Edvardsson \etal 1993), and assuming that this relation applies to open
clusters
as well, the iron gradient corresponds to
an oxygen gradient $\Delta$log(O/H)/$\Delta$R = -0.06, flatter indeed than
that derived from HII regions and more similar to that of objects (as the
PNeII discussed above) a few Gyr old.

\begin{figure}
  \vspace{5.5truecm}
  \caption{Radial distribution of open clusters metallicity as derived from
  Friel \& Janes (1993) and Thogersen \etal (1993) samples. The clusters have
  been divided in age bins and the linear best fit for each bin is shown
  (dotted line for the oldest bin, long-dashed for the 4-5 Gyr bin,
short-dashed
  for the 2-4 Gyr bin, and solid line for the youngest one.}
\end{figure}

The most striking feature of Friel \& Janes' sample is that at each galactic
radius the oldest clusters are also the most metal rich (see Fig.5). It is true
that the published clusters in the oldest age bin ($\geq$8 Gyr) are only four
and the corresponding statistics is therefore too poor; however
two additional clusters of roughly the same age have been found (Friel 1994,
private communication) with less extreme metal abundances but still higher
than average. It is of crucial importance to verify this result
with a larger sample of old clusters and with more accurate and
homogeneous methods to derive their ages, chemical abundances and galactic
original locations. It might well be, in fact, that this anomaly is fictitious
and resulting from the uncertainty in the metallicity and/or, more probably,
age determination. However, if confirmed, this phenomenon would have remarkable
implications on our understanding of the Galaxy evolution, because it is
opposite to any intuitive age-metallicity relation derivable from {\it
steady-state} scenarios where old stars are inevitably more metal poor than
young objects and may result from short, intense phenomena not considered in
our models.

Another characteristics of old open clusters is that all of them are located
beyond 7-7.5 kpc from the galactic center, contrary to younger clusters
which are likely to be equally distributed everywhere in the disk (Janes
\& Phelps 1994). On the one hand, the external location of the older clusters
in the observed sample
might not reflect an odd distribution of all the clusters formed several Gyr
ago and be the result of a more efficient disruption in the inner than in the
outer regions. On the other hand, it may instead correspond to a non
homogeneous
star formation activity, perhaps related to external phenomena like the first
impact on the disk of the Magellanic Stream (Sofue 1994). The latter
scenario might also provide an explanation to the large metallicity of the
oldest clusters, in terms of a transitory metal enhancement of the ISM due
to the larger SFR triggered by the sudden event and later smeared out during
the following {\it steady-state} evolution.

\section[]{Summary}

In conclusion, the comparison between the abundance distributions in the
galactic disk predicted by {\it standard} chemical evolution models and
derived from observational data can be summarized as follows:\\
a) Abundance distributions derived from observations of gaseous objects of
various ages (HII regions and PNe) are very well reproduced by {\it
steady-state} models with slowly decreasing SFR and large - long lasting
infall of metal poor gas.\\
b) Average abundance distributions derived from stars are also reproduced.
These {\it standard} models, however, do not reproduce the observed spread
in the metallicity distribution of field stars and the anomalously high
[Fe/H] of the oldest open clusters.

We must bear in mind, however, that old stars can have quite eccentric orbits
and can therefore have formed in galactic regions different from those where
they are observed now. According to Fran\c cois \& Matteucci this might explain
most of the abundance spread in the Edvardsson's \etal sample. On the other
hand, Carraro \& Chiosi (1994) have suggested that the same argument cannot
apply to the case of old open clusters, for which other explanations are
thus needed, unless all the inconsistencies with the model predictions can
be attributed to observational errors.

A possible reason for the different agreement found for gas and stellar
objects is that the gas mixes rapidly, compared to the timescales for
galactic evolution, therefore forgets local perturbations occurred in the
past and follows the {\it steady-state} scenario. Stars, instead, keep
memory of the local perturbations occurring at, or just before, their birth
and therefore deviate more from that scenario, showing intrinsic large
scatter and anomalous behaviours. To interpret in detail their observed
features more sophisticated models taking into account also the small scale,
short term phenomena are therefore required.

\vskip 1truecm
I wish to thank Francesca Matteucci for always being ready to discuss and
compare the results and the different approaches of our models: a
praiseworthy attitude rather unusual among theoreticians.

\vskip 2truecm
\centerline{\bf Discussion}
\vskip 0.5truecm
{\bf Palous}. As you have shown, SFR(R) is a function of both $\sigma_{gas}$
and
$\sigma_{tot}$: SFR = $\psi (\sigma_{gas},\sigma_{tot})$ $e^{-t/\tau}$. In the
model of propagating SF, recently published by Palous, Tenorio-Tagle \& Franco
(1994, M.N.R.A.S.) the SFR $\propto [\sigma_{gas}, sheer(R)]$. Sheer follows
the total mass distribution and is decreasing with R. It seems then that the
propagating SF model reaches conclusions similar to yours.

{\bf Chernin}. Do I understand you correctly that there might be two
characteristic time scales of the chemical evolution of the galactic disk,
15 and 5 Gyr, and that the available observational data do not enable us to
make a choice between them ? And, by the way, is there any room for dark
matter in your models, whatever its physical nature may be ?

{\bf Tosi}. Yes, with a proper combination of the other parameters, both a
SFR with e-folding time 15 Gyr and one with 5 Gyr provide results consistent
with the disk observational data. I haven't included dark matter in my models.

{\bf Tenorio-Tagle}. Would you comment about outflows ?

{\bf Tosi}. Outflows haven't been introduced yet in {\it standard} chemical
evolution models, because we don't have enough observational constraints on
their chemical composition, on the significance of the phenomenon or on the
final fate of the ejected gas (does it eventually escape from the system or
falls back, and where ?). In absence of such information, too many free
parameters should be assumed in the models to allow for a safe interpretation
of the results.

{\bf Serrano}. What would be the effect on the standard model of outflows
larger at small R and smaller at large R ?

{\bf Tosi}. If the ejected gas is lost for ever,
it should lead to a reduction of the inner abundances and therefore
to a flattening of the radial gradient of the outflowing chemical elements. If
the wind is made only of the gas ejected by SNe, the effect will be restricted
to the elements, like oxygen, synthesized by their progenitors, otherwise
it will be on all the metals.

{\bf Martin}. In your models you need a total amount of infalling gas mass
comparable to the mass of the disk. So much infall cannot come from the
Magellanic Clouds: where from do you think it's coming ?

{\bf Tosi}. The infalling gas has presumably a composite origin: at early
epochs
it must have been mostly collapsing halo material, with a small component
of extragalactic gas; nowadays it is most probably due to external gas. The
sum of these components can easily account for the required
$\sim 10^{10}$M$_{\odot}$.

{\bf Andersen}. I am still somewhat concerned about the amount and age of
data on which some crucial features of the models depend rather critically.
Edvardsson \etal did not discuss possible radial gradients from their data,
simply because we did not have a fair sample of the stars originally inside
or outside the solar circle.

{\bf Tosi}. I mentioned indeed that the statistics on the older and more
distant
bins is quite poor. With this caveat in mind, I think however that it is
useful to check whether or not the model predictions are consistent
with such a large and good sample of data.

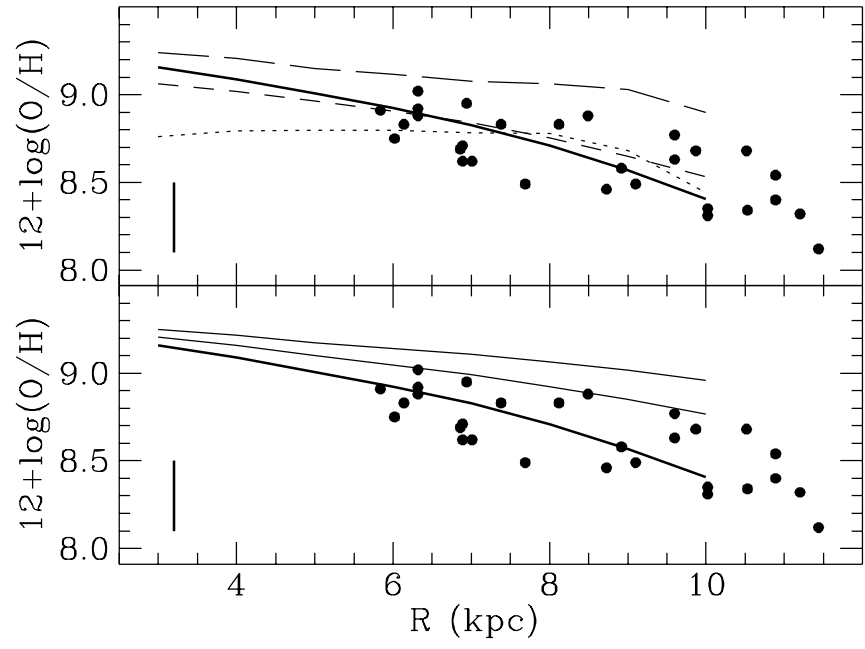
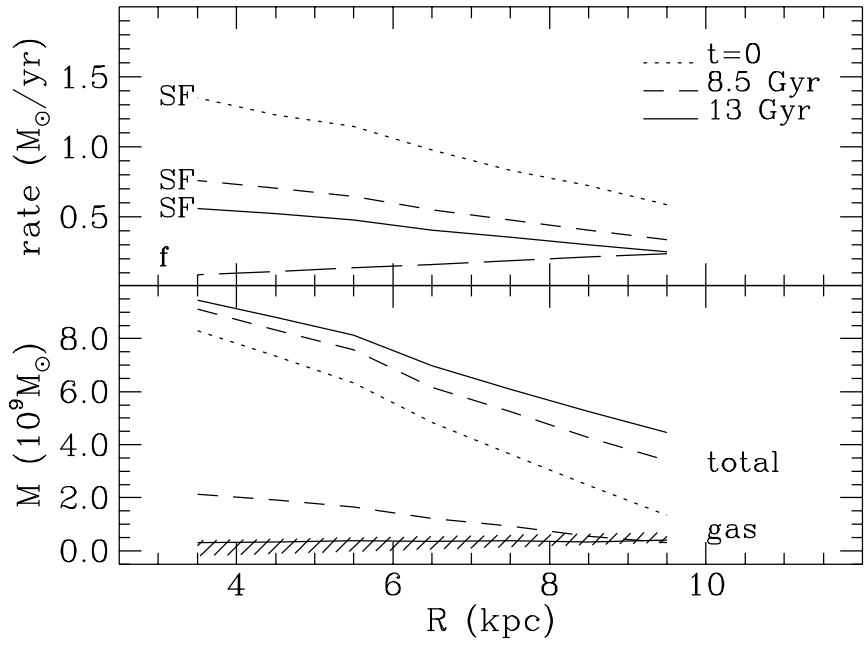
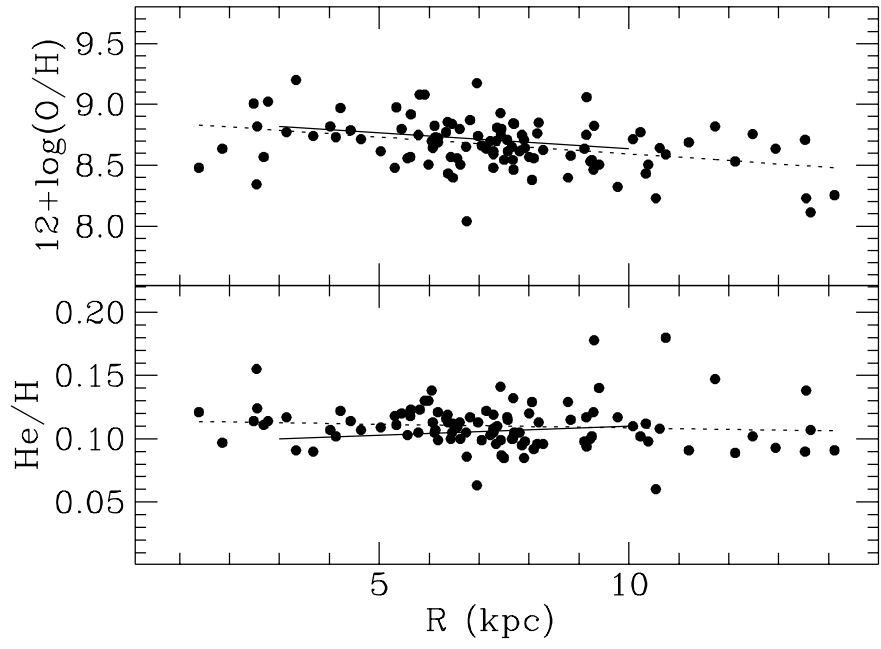
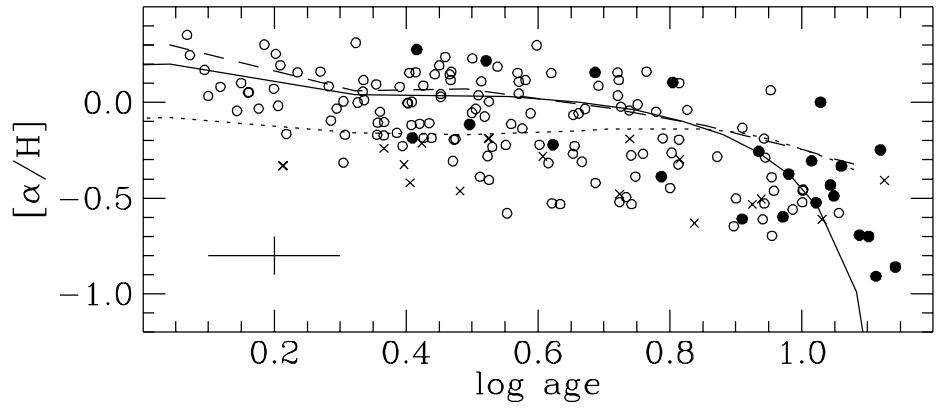
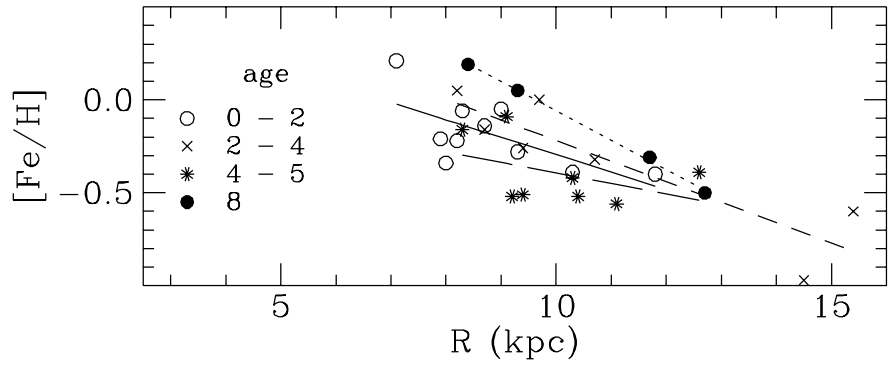
\end{document}